\documentstyle{article}
\begin{document}
\bibliographystyle{unsrt}
\title{Renormalization by Projection:\\
On the Equivalence of the Bloch-Feshbach Formalism and Wilson's
Renormalization}
\author{Jochen M\"uller \\
Max-Planck-Institut f\"ur Kernphysik \\
Postfach 103980, 69029 Heidelberg, Germany
\and 
Jochen Rau\thanks{Address after October 1, 1996:
Max-Planck-Institut f\"ur Physik komplexer Systeme,
Bayreuther Stra{\ss}e 40 Haus 16,
01187 Dresden, Germany.} \\
European Centre for Theoretical Studies \\
in Nuclear Physics and Related Areas (ECT$^*$) \\
Villa Tambosi \\
Strada delle Tabarelle 286, 38050 Villazzano (Trento), Italy}
\date{Revised version, \today}
\maketitle
\begin{abstract}
We employ projection operator techniques in Hilbert space to derive
a continuous sequence of effective Hamiltonians which describe the
dynamics on successively larger length scales.
We show for the case of $\phi^4$ theory that the masses and
couplings in these effective Hamiltonians vary in accordance
with 1-loop renormalization group equations.
This is evidence for an intimate connection between
Wilson's renormalization and the venerable Bloch-Feshbach formalism.
\end{abstract}
{PACS numbers {11.10.Gh, 11.10.Hi, 64.60.Ak, 21.30.+y}}
%\\\\\\
%ECT*/Nov/95-04
%
\clearpage
\section{Introduction}
The Bloch-Feshbach formalism, developed in the late 50's to
describe selected features of nuclear dynamics
\cite{bloch,haw},
and Wilson's renormalization group, developed in the early 70's
to better understand critical phenomena
\cite{kadanoff,wilson,wegner,polchinski},
are based on the
same physical idea:
eliminate irrelevant modes in order to focus on the
dynamics of few selected degrees of freedom.
The Bloch-Feshbach formalism employs projection operators in Hilbert space
in order to determine the effective Hamiltonian in some
restricted model space, thereby discarding dynamical information
that pertains to the irrelevant modes.
In a similar spirit, renormalization --as originally conceived in the context
of statistical physics--
is a mathematical tool that allows one to 
iteratively eliminate short-wavelength modes and thus to arrive at
effective (``renormalized'')
theories which describe the dynamics on successively
larger length scales.
In both cases the irrelevant modes no longer appear explicitly
in the effective theory, but their residual influence on the
dynamics of the remaining modes is taken into account through
adjustments of the effective interaction.
The power and elegance of both methods
derives from the fact that they thus allow one to study 
accurately selected features
of the dynamics, such as its infrared limit, 
without ever having to solve the
full underlying microscopic theory.
 
That renormalization and the old projection technique of the 
Bloch-Feshbach formalism
are in fact closely
related, and that in some cases the former can be regarded as
a special case of the latter,
has already been hinted at by Anderson in his 
``poor man's scaling'' approach to the Kondo problem \cite{anderson},
by Seke in his projection-method treatment of the nonrelativistic
Lamb shift \cite{seke},
and by more recent studies of a simple quantum
mechanical model \cite{fields};
and it is clearly suggested by the modern view of renormalization as
yielding a continuous sequence of
effective theories
\cite{wilson,wegner,polchinski}.
In our letter we wish to supply further evidence for this connection by
calculating the 1-loop renormalization of $\phi^4$ theory
with the help of the old Bloch-Feshbach techniques.
  
We first introduce the basic mathematical framework.
Let $H$ denote the original (full) Hamiltonian,
$P$ a projection operator which projects the original Hilbert space
onto some selected subspace, and $Q=1-P$ its complement.
For $\phi^4$ theory, as for any
many-particle theory,
the elimination of short-wavelength modes corresponds 
to a projection in Fock space:
lowering the momentum
cutoff from some original value $\Lambda$ to
\begin{equation}
\Lambda(\Delta s):=\exp(-\Delta s)\Lambda
\quad,\quad \Delta s\ge 0
\quad,
\end{equation}
is effected by a
projection operator $P(\Delta s)$
which acts on 
$n$-particle states 
according to 
\begin{equation}
P(\Delta s)|{k}_1\ldots {k}_n\rangle:=
\prod_{i=1}^n \theta(\Lambda-e^{\Delta s} |{k}_i|)
|{k}_1\ldots {k}_n\rangle
\quad,
\end{equation}
where $\{{k}_1\ldots {k}_n\}$ denote the particle 
momenta\footnote{In a relativistic theory this definition of $P$
is not covariant.}.
Provided 
the eliminated modes ($Q$) have energies on
some large characteristic scale $\omega_\Lambda$,
much larger than the energy scale at which we want to
study the system's physical properties,
then 
in the reduced Hilbert space the dynamics is 
approximately\footnote{up to corrections of order $O(E/\omega_\Lambda)$,
with $E$ being the (low) physical scale} 
governed by 
the effective Hamiltonian \cite{bloch,haw}
\begin{equation}
H_{\rm eff}(\Delta s) \approx P(\Delta s)HP(\Delta s) + 
\Sigma(\Delta s)
\label{proj_form}
\end{equation}
with
\begin{equation}
\Sigma(\Delta s):= -P(\Delta s)HQ(\Delta s)
{1\over Q(\Delta s)HQ(\Delta s)}Q(\Delta s)HP(\Delta s)
\quad.
\end{equation}
If the original Hamiltonian can be decomposed into a free
and an interaction part,
$H=H^{(0)}+V$,
where the free part $H^{(0)}$ commutes with the
projection, 
then to lowest nontrivial (i. e., second) order perturbation theory
\begin{equation}
\label{approx_eff}
\Sigma(\Delta s)\approx -P(\Delta s)V{Q(\Delta s)\over H^{(0)}}VP(\Delta s)
\quad.
\end{equation}
It is this approximation formula which we shall use
in our subsequent
calculations.
As the parameter $\Delta s$ increases,
the masses and coupling constants in $H_{\rm eff}(\Delta s)$
vary.
We claim that this flow is equivalent to a conventional
renormalization.
In particular, we claim that
the above approximation formula 
yields the renormalization of
$\phi^4$ theory in agreement
with diagrammatic 1-loop calculations.
We will show that
the projection $PHP$ of the Hamiltonian is responsible for
the renormalization of the mass, while
the additional term $\Sigma$ gives rise to the renormalization of the
coupling constant.

The $\phi^4$ Hamiltonian describes coupled anharmonic
oscillators in spatial dimension $(d-1)$.
It reads
\begin{eqnarray}
H&=&
{1\over2}\int d^{d-1}x :\!\left[\pi(x)^2
+ |\nabla\phi(x)|^2 + m^2\phi(x)^2\right]\!:
+g {\mu^\epsilon\over 4!} \int d^{d-1}x\,\phi(x)^4
\nonumber \\
&=:& 
H^{(0)}[\pi,\phi] + V[\phi]
\quad,
\end{eqnarray}
where $\epsilon:=(4-d)$, $\mu$ denotes
a reference momentum scale of the interaction, 
$m$ the mass, $g$ the coupling constant,
and $:\![\ldots]\!:$ means normal
ordering.
The field $\phi$ and its conjugate momentum $\pi$ 
are time-independent (Schr\"odinger picture) operators which satisfy the
commutation relations for bosons.
They can be expressed in terms of annihilation and creation
operators $a$, $a^\dagger$; e. g.,
\begin{equation}
\phi(x)=\int_{|k|\le\Lambda} {d^{d-1}k\over (2\pi)^{d-1}}\,
{1\over\sqrt{2\omega_k}}(a_k+a^\dagger_{-k})\cdot \exp(ikx)
\end{equation}
with
\begin{equation}
[a_k,a^\dagger_q]=(2\pi)^{d-1}\delta^{d-1}(k-q)
\end{equation}
and
\begin{equation}
\omega_k:=\sqrt{k^2+m^2}
\quad.
\end{equation}
In our subsequent calculations it will prove useful
to decompose the field operators according to
\begin{equation}
\phi(x)=\phi_<(x) + \phi_>(x)
\quad,
\end{equation}
where $\phi_<$ contains all momentum modes up to the
lower cutoff $e^{-\Delta s}\Lambda$, whereas $\phi_>$
contains the remaining modes in the
``shell'' $[e^{-\Delta s}\Lambda,\Lambda]$;
analogously for $\pi=\pi_<+\pi_>$.
Any power of $\phi$ can then be written as a polynomial
\begin{equation}
\phi^n=\sum_{m=0}^n \left(
{n\atop m}
\right)
\phi_>^{n-m} \phi_<^m 
\quad,
\end{equation}
the binomial coefficients $(:)$ counting the number of ways
in which the $\phi_<$ and $\phi_>$ can be arranged.
Since the slow field operators $\phi_<$ act 
in the reduced Hilbert space only, they commute with
the projection,
\begin{equation}
[\phi_<,P(\Delta s)]=0
\quad.
\end{equation}
The fast field operators $\phi_>$, on the other hand, 
do not change the particle content within the reduced Hilbert
space, and hence for an arbitrary polynomial $f(\phi_>)$
it is
\begin{equation}
P(\Delta s)f(\phi_>)P(\Delta s)= 
\langle 0|f(\phi_>)|0\rangle\cdot P(\Delta s)
\quad,
\label{replace}
\end{equation}
where $|0\rangle$ denotes the vacuum.
With the help of Wick's theorem all such vacuum expectation
values can be reduced to sums and products of
\begin{equation}
\langle 0|\phi_>(x)\phi_>(y)|0\rangle=
\int_{\rm shell}\!{d^{d-1}k\over (2\pi)^{d-1}}
\,{1\over 2\omega_k}\exp[ik\cdot(x-y)]
\quad.
\end{equation}
\section{Renormalization Group Equations}
We first show that the projection $PHP$ of the Hamiltonian gives rise
to a renormalization of the mass.
The unperturbed part of $PHP$ simply contributes
\begin{equation}
{P(\Delta s)H^{(0)}[\pi,\phi]P(\Delta s)}
=
H^{(0)}[\pi_<,\phi_<]
\end{equation}
(up to an additive constant) and thus does not entail 
any modification of the mass or coupling constant.
The interaction, on the other hand, yields
\begin{equation}
{P(\Delta s)V[\phi]P(\Delta s)}
={g}{\mu^\epsilon\over 4!}
\sum_{n=0}^4 \left({4\atop n}\right)
\int\! d^{d-1}x \,\langle 0|\phi_>(x)^{4-n}|0\rangle
\,\phi_<(x)^n
\quad.
\end{equation}
Odd powers of $\phi_>$ have a vanishing vacuum expectation
value, so only the terms with $n=0,2,4$ survive.
They contribute to the zero-point (vacuum) energy, mass
term and $\phi^4$ interaction term,
respectively, of the effective Hamiltonian.
The latter contribution is
just the original $\phi^4$ interaction term
restricted to the slow modes, with no change in
the associated coupling $g$.
In contrast, the contribution to the mass term leads to a nontrivial
modification
\begin{equation}
{1\over2}\Delta m^2=
{g}{\mu^\epsilon\over 4!}
\left({4\atop 2} \right)
\langle 0|\phi_>(x)^2|0\rangle
\quad.
\end{equation}
Taking the flow parameter $\Delta s$ to be infinitesimal,
the vacuum expectation value is given by
\begin{equation}
\langle 0|\phi_>(x)^2|0\rangle
={S_{d-1}\over 2(2\pi)^{d-1}}\Lambda^{d-2}
{\Lambda\over\omega_\Lambda}
\Delta s
\quad,
\end{equation}
where $S_{d-1}$ denotes the surface of a unit shell in
$(d-1)$-dimensional momentum space.
Expanding
\begin{equation}
{\Lambda\over\omega_\Lambda}
= 1-{m^2\over 2\Lambda^2}+O(m^4/\Lambda^4)
\end{equation}
we obtain
\begin{equation}
\Delta m^2
\approx -{S_{d-1}g\over 8(2\pi)^{d-1}}
\left({\mu\over\Lambda}\right)^\epsilon 
(m^2-2\Lambda^{2})
\Delta s
\quad.
\end{equation}
For $d=4$ and a spherical cut in 3-momentum space
($S_{d-1}=4\pi$) this yields
\begin{equation}
\Delta m^2
= -{g\over 16\pi^2}
(m^2-2\Lambda^2)
\Delta s
\quad,
\end{equation}
in agreement with well-known 1-loop results
\cite{fisher}\footnote{The numerical factor 
in front of the nonuniversal
$\Lambda^2$-term may differ. This is due to
different definitions of the cutoff in 3- or 4-momentum
space, respectively.
The negative overall 
sign stems from our definition of
the flow parameter: for $\Delta s>0$ we are
{\em lowering} the cutoff,
$d/ds=-\Lambda\,d/d\Lambda$.}.

Next we show that the additional term $\Sigma$ in the effective
Hamiltonian yields
a renormalization of the coupling constant.
From the approximate expression (\ref{approx_eff})
we read off that $\Sigma$
contains terms of order 
$\phi_<^0$, $\phi_<^2$, $\phi_<^4$ and $\phi_<^6$, 
contributing to vacuum energy, mass term, $\phi^4$ interaction term
and a new $\phi^6$ interaction term, respectively, of
the effective Hamiltonian.
We denote the various contributions by $\Sigma_n$,
$n\in\{0,2,4,6\}$.
It is
\begin{eqnarray}
\Sigma_n(\Delta s)
&=&
-\left[{g}{\mu^\epsilon\over 4!}\right]^2
\sum_{m=\max\{0,n-3\}}^{\min\{3,n\}} 
\left({4\atop m}\right)\left({4\atop {n-m}}\right)
\nonumber \\
&& \times
\int\! d^{d-1}x \int\! d^{d-1}y\,
\phi_<(x)^m  \phi_<(y)^{n-m}
\nonumber \\
&&\times
\left[P\phi_>(x)^{4-m}{Q\over H^{(0)}}
\phi_>(y)^{4+m-n} P
\right]
\quad.
\label{self_n}
\end{eqnarray}
The contribution $\Sigma_2$ to the mass term
contains vacuum expectation values 
of the form 
$\langle 0|\phi_>^2 Q\phi_>^4|0\rangle$,
$\langle 0|\phi_>^3 Q\phi_>^3|0\rangle$ and
$\langle 0|\phi_>^4 Q\phi_>^2|0\rangle$, respectively,
which are all at least of order $(\Delta s)^2$
and thus negligible for infinitesimal values of $\Delta s$.
The contribution $\Sigma_6$ to the new
$\phi^6$ interaction term, on the other hand, scales as
\begin{eqnarray}
\Sigma_6
&\sim&
\int\! d^{d-1}x \!\int\! d^{d-1}y\, \phi_<(x)^3
\phi_<(y)^3
\langle 0|\phi_>{Q\over H^{(0)}}\phi_>|0\rangle
\nonumber \\
&\sim&
{1\over \omega_\Lambda^2}\int\! d^{d-1}x\, \phi_<(x)^6
\end{eqnarray}
and is thus suppressed for large values of the cutoff.
Within our approximations, therefore, the only contribution
which could significantly alter the effective Hamiltonian
is $\Sigma_4$;
it could modify the
$\phi^4$ interaction and hence the associated coupling 
constant $g$.

For $n=4$ the sum over $m$ in Eq. (\ref{self_n}) runs from $1$ to $3$;
a non-vanishing contribution stemming, however, only
from $m=2$.
In the remaining term we replace, according to Eq. (\ref{replace}),
\begin{equation}
P\phi_>(x)^2{Q\over H^{(0)}}\phi_>(y)^2P
=
\langle 0|\phi_>(x)^2 Q \phi_>(y)^2|0\rangle\cdot
{P\over 2\omega_\Lambda +H^{(0)}}
\quad;
\end{equation}
where the vacuum expectation value can be further reduced to
\begin{equation}
\langle 0|\phi_>(x)^2Q\phi_>(y)^2|0\rangle
= 
2\langle 0|\phi_>(x)\phi_>(y)|0\rangle^2
\quad.
\end{equation}
Provided the eliminated shell in momentum space is invariant
under time reversal ($k\to -k$),
the right-hand side is real and
positive definite. It is a distribution in $(x-y)$
whose width scales as $1/\Lambda$.
For large cutoff, therefore, it can be
approximated by
\begin{eqnarray}
2\langle 0|\phi_>(x)\phi_>(y)|0\rangle^2
&\approx&
2\langle 0|\phi_>(x)^2|0\rangle
\cdot {1\over 2\omega_\Lambda}
\delta^{d-1}(x-y)
\nonumber \\
&=&
{S_{d-1}\over 2(2\pi)^{d-1}\Lambda^{\epsilon-1}}
\delta^{d-1}(x-y)\cdot\Delta s
+O(m^2/\Lambda^2)
\quad.
\end{eqnarray}
As long as the mass $m$ and the energy 
$H^{(0)}$ of the external, slow modes
are negligible compared to the cutoff $\Lambda$,
we immediately find
\begin{equation}
\Sigma_4(\Delta s)
=-{g}{\mu^\epsilon\over 4!}
\left({4\atop2}\right)^2
{S_{d-1}\over 4(2\pi)^{d-1}\Lambda^\epsilon}
\Delta s\cdot V[\phi_<]
\quad.
\end{equation}
The resultant modification of the coupling constant reads
\begin{equation}
\Delta g=
-{3S_{d-1} g^2\over 8(2\pi)^{d-1}}
\left({\mu\over\Lambda}\right)^\epsilon
\Delta s
\quad;
\end{equation}
which for $d=4$ and a spherical cut in 3-momentum space 
reduces to
\begin{equation}
\Delta g
=-{3g^2\over 16\pi^2}\Delta s
\quad,
\end{equation}
again in agreement with known 1-loop results
\cite{fisher}.
\section{Discussion}
Renormalization is equivalent to determining a continuous sequence of
effective Hamiltonians in smaller and smaller
Hilbert spaces, obtained by successive elimination
of short-wavelength modes.
For the case of $\phi^4$ theory we have shown that,
to 1-loop order,
these effective Hamiltonians
and hence the renormalization flow of the masses and couplings
can be determined with the
help of the venerable Bloch-Feshbach formalism.
This finding might be interesting for several
reasons:
\begin{enumerate}
\item
Renormalization is often formulated in terms of 
functional integrals and diagrams,
while projection techniques in Hilbert space are
based on algebraic concepts such as
linear subspaces and operators. 
Building a bridge between these different languages offers
a new and interesting conceptual perspective and potentially
broadens the range of available
calculational tools.
\item
Projection techniques permit the elimination not just of
short-distance information, but also of
other kinds of information deemed irrelevant,
such as high angular momenta, spin degrees of freedom,
or entire particle species.
Adopting the projection approach may therefore open the
way to new, more general renormalization schemes.
\item
Finally, projection techniques provide
a common framework for both renormalization and the transition
to macroscopic transport theories \cite{rau}.
They are thus a natural language to 
study issues such as the renormalization of
macroscopic transport
equations, or effective kinetic theory \cite{jeon}.
\end{enumerate}
\section*{Acknowledgement}
We thank S. Kehrein, J. Polonyi, J. P. Vary,
F. J. Wegner and H. A. Weidenm\"uller for
helpful discussions.
JR acknowledges
financial support by the Heidelberger Akademie der Wissenschaften
and by the European Union HCM fellowship programme.

\end{document}